\documentclass{bioinfo}
\copyrightyear{2018} \pubyear{2018}

\access{Publication Date: March 05, 2018}
\appnotes{New Tool Release}

\begin{document}
\firstpage{1}

\subtitle{New Tool Release}

\title[Python for SEAL]{PySEAL: A Python wrapper implementation of the SEAL homomorphic encryption library}
\author[Sample \textit{et~al}.]{Alexander J. Titus\,$^{\text{\sfb 1,2,4}}$, Shashwat Kishore\,$^{\text{\sfb 3,4}}$, Todd Stavish\,$^{\text{\sfb 3}}$, Stephanie M. Rogers\,$^{\text{\sfb 1}}$, and Karl Ni\,$^{\text{\sfb 3,}*}$}

\address{$^{\text{\sf 1}}$B.Next, IQT Labs, Arlington, VA, 22201, USA \\
$^{\text{\sf 2}}$Quantitative Biomedical Sciences, Dartmouth School of Graduate and Advanced Studies, Hanover, NH, 03755, USA \\
$^{\text{\sf 3}}$Lab41, IQT Labs, Menlo Park, CA, 94025, USA\\
$^{\text{\sf 4}}$Authors contributed equally}

\corresp{$^\ast$To whom correspondence should be addressed.}

\history{Published on March 05, 2018}

\editor{Primary Authors: AJ Titus \& S Kishore}

\abstract{\textbf{Motivation:} The ability to perform operations on encrypted data has a growing number of applications in bioinformatics, with implications for data privacy in health care and biosecurity. The \textit{SEAL} library is a popular  implementation of fully homomorphic encryption developed in C++ by Microsoft Research. Despite the advantages of C++, Python is a flexible and dominant programming language that enables rapid prototyping of bioinformatics pipelines.\\
\textbf{Results:} In an effort to make homomorphic encryption accessible to a broader range of bioinformatics scientists and applications, we present a Python binding implementation of the popular homomorphic encryption library, \textit{SEAL}, using \emph{pybind11}. The software contains a Docker image to facilitate easy installation and execution of the SEAL build process.\\
\textbf{Availability:} All code is publicly available at \textit{https://github.com/Lab41/PySEAL}\\
\textbf{Contact:} \href{lab41@iqt.org}{lab41@iqt.org}\\
\textbf{Supplementary information:} Supplementary information is available on the Lab41 GitHub.}

\maketitle

\section{Introduction}

The volume of available genomic data is growing rapidly as the cost of sequencing continues to decline. Biomedical research advances can greatly benefit from this wealth of data, but the sensitive nature of such information demands that privacy and security concerns be given full attention. For example, recent work has demonstrated an ability to re-identify individuals using their genomic data (\citealp{Gymrek2013,Lippert2017}). Furthermore, many analyses, such as the Genome Wide Association Studies (GWAS), require large sample sizes that are often difficult to collect at a single clinical site, which highlights the challenge of maintaining the security of sensitive data being collected and shared among multiple platforms at multiple geographic locations.

In 2010, the Center for Integrating Data for Analysis, Anonymization and SHaring (iDASH) was founded at the University of California, San Diego to address privacy challenges to research progress. At the annual iDASH Privacy \& Security Workshop, teams from around the world develop novel methods for private and secure computation on health data, and extend the applications of homomorphic encryption into the areas of biological data storage, transport, and computation (\citealp{Kim2015}).

\begin{figure}[!tpb]
\centerline{\includegraphics[width=\columnwidth]{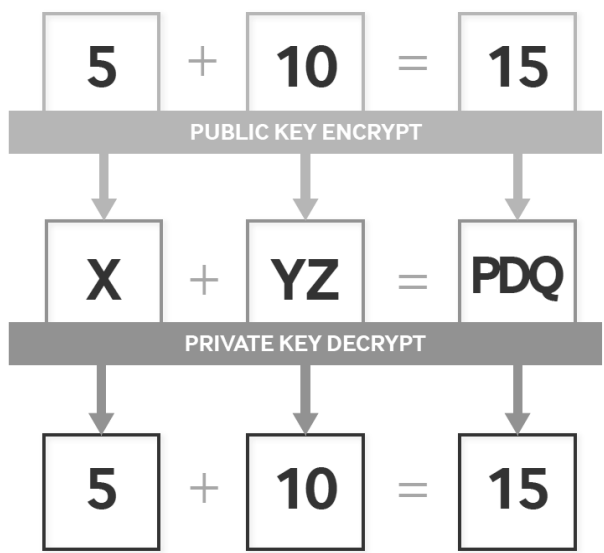}}
\caption{In a fully homomorphic encryption scheme, operations on ciphertext space are mirrored in the plaintext space after decryption.}\label{fig:01}
\end{figure}

Homomorphic Encryption offers a promising approach to addressing privacy and security concerns in the handling of sensitive genomic data, because it allows data to remain encrypted even during use, and it offers a security method to enable comparisons across disparate hardware and geographic distances.  Under a fully homomorphic encryption cryptoscheme (\citealp{Gentry2009}), any arbitrary set of operations that take place on the ciphertext (encrypted space) are mirrored in unencrypted space (Figure \ref{fig:01}). 

Here, we present a Python binding of the Simple Encrypted Arithmetic Library (\textit{SEAL}) homomorphic encryption library (\citealp{SEALman}). \textit{SEAL} is maintained by the Cryptography Research Group at Microsoft Research and implements a fully homomorphic encryption scheme in C++. The software has strong applications in bioinformatics (\citealp{Dowlin2017}). While C++ is appropriate for software engineering and developed deployments on established architectures, biologists and bioinformatics scientists,  especially those with less-than-formal computer science training, often prefer rapid prototyping scripting languages like Python. Therefore, we present \textit{PySEAL}, a Python wrapper implementation of the \textit{SEAL} library using \textit{pybind11}. The \textit{PySEAL} package implements \textit{SEALv2.3} and the package is under active maintenance.

\begin{methods}
\section{Methods}
\textit{PySEAL} is a Python wrapper for the Simple Encrypted Arithmetic Library (\textit{SEAL}) homomorphic encryption library. The \textit{PySEAL} software is implemented using \textit{pybind11} and is released as a Docker container, a C++ \textit{SEAL} example build, and a Python wrapper build. The software also includes bash scripts to build the Docker container and execute a series of examples using \textit{PySEAL} functions. 

The \textit{PySEAL} code is released on the Lab41 GitHub page (https://github.com/Lab41/PySEAL) and is under active maintenance. The README includes an overview description of the steps to instantiate and use the fully homomorphic encryption scheme. 

Briefly, the first step to using the library for basic encryption tasks is to instantiate a new \verb"EncryptionParameters" object and set its modulus attributes. These parameters are stored in a \verb"SEALContext" object which checks the validity of the parameters. Once validated, these parameters are used to create encryption keys and \verb"Encryptor" and \verb"Decryptor" objects. 

For operations, \verb"IntegerEncoder" and \verb"Evaluator" objects are included to encrypt input data and perform operations on the ciphertext. \textit{SEALv2.3} includes functionality to compute operations between ciphertext and plaintext, allowing for runtime improvements due to reduced encryption operational overhead. 

\end{methods}

\section{Applications}
The combination of such operations has enabled encrypted mutation searching (\citealp{Çetin2017}) and encrypted bloom filter searching for sequence-to-sequence comparisons. Other recent bioinformatics applications of homomorphic encryption schemes include the computation of edit distances (\citealp{Cheon2015}), goodness of fit testing (\citealp{Lauter2014}), linkage disequilibrium measurement (\citealp{Lauter2014}), minor allele frequency computations (\citealp{Kim2015}), and  chi-square statistic computations (\citealp{Kim2015}). 

The homomorphic encryption package has many applications outside genomics as well. For example, homomorphic encryption has recently been used in conjunction with other methods to implement biometric authentication schemes (\citealp{Karabat2015}) and secure E-commerce recommendation systems (\citealp{Ahila2015}). In general, homomorphic encryption could potentially be used for any of a host of tasks that involve multiple parties sharing sensitive data.

\section{Conclusion}
Homomorphic encryption is a promising application for secure computation in biomedical research and clinical data applications. The C++ \textit{SEAL} software from Microsoft Research is under active development for performance improvements, as homomorphic encryption algorithm runtime is a limiting factor to deploying homomorphic encryption in production systems. With \textit{PySEAL}, we reduce the challenge of developing in C++ and provide the bioinformatics research community with a tool to quickly prototype and iterate on new methods, allowing performance and implementations to progress in parallel.\\

\section*{Acknowledgements}
The authors thank Lisa Porter for her continued support and guidance. This work was supported in part by a Big Data to Knowledge (BD2K) training grant to AJT (T32LM012204).

% \section*{Funding}

% This work has been supported by...\\

\bibliographystyle{natbib}

\bibliography{pySEAL}

\begin{thebibliography}{}

\bibitem[Ahila and Shunmuganathan(2015)Ahila and Shunmuganathan]{Ahila2015}
Ahila, S.~S. and Shunmuganathan, K.~L. (2015).
\newblock Role of agent technology in web usage mining: Homomorphic encryption
  based recommendation for e-commerce applications.
\newblock {\em Wireless Personal Communications\/}.

\bibitem[{\c{C}}etin {\em et~al.}(2017){\c{C}}etin, Chen, Laine, Lauter,
  Rindal, and Xia]{Çetin2017}
{\c{C}}etin, G.~S., Chen, H., Laine, K., Lauter, K., Rindal, P., and Xia, Y.
  (2017).
\newblock Private queries on encrypted genomic data.
\newblock {\em BMC Medical Genomics\/}, {\bf 10}(2), 45.

\bibitem[Chen and Laine(2017)Chen and Laine]{SEALman}
Chen, H. and Laine, K. (2017).
\newblock {Simple Encrypted Arithmetic Library - SEAL v2.2}.
\newblock Technical report.

\bibitem[Cheon {\em et~al.}(2015)Cheon, Kim, and Lauter]{Cheon2015}
Cheon, J.~H., Kim, M., and Lauter, K. (2015).
\newblock Homomorphic computation of edit distance.
\newblock {\em Financial Cryptography and Data Security\/}.

\bibitem[Dowlin {\em et~al.}(2017)Dowlin, Gilad-Bachrach, Laine, Lauter,
  Naehrig, and Wernsing]{Dowlin2017}
Dowlin, N., Gilad-Bachrach, R., Laine, K., Lauter, K., Naehrig, M., and
  Wernsing, J. (2017).
\newblock {Manual for using homomorphic encryption for bioinformatics}.
\newblock {\em Proceedings of the IEEE\/}, {\bf 105}(3), 552--567.

\bibitem[Gentry(2009)Gentry]{Gentry2009}
Gentry, C. (2009).
\newblock {Fully homomorphic encryption using ideal lattices.}
\newblock In {\em STOC\/}, volume~9, pages 169--178.

\bibitem[Gymrek {\em et~al.}(2013)Gymrek, McGuire, Golan, Halperin, and
  Erlich]{Gymrek2013}
Gymrek, M., McGuire, A.~L., Golan, D., Halperin, E., and Erlich, Y. (2013).
\newblock {Identifying Personal Genomes by Surname Inference}.
\newblock {\em Science\/}, {\bf 339}(6117), 321 -- 324.

\bibitem[Karabat {\em et~al.}(2015)Karabat, Mehmet, Erdogan, and
  Savas]{Karabat2015}
Karabat, C., Mehmet, S.~K., Erdogan, H., and Savas, E. (2015).
\newblock Thrive: threshold homomorphic encryption based secure and privacy
  preserving biometric verification system.
\newblock {\em EURASIP Journal on Advances in Signal Processing\/}.

\bibitem[Kim and Lauter(2015)Kim and Lauter]{Kim2015}
Kim, M. and Lauter, K. (2015).
\newblock {Private genome analysis through homomorphic encryption}.
\newblock {\em BMC medical informatics and decision making\/}, {\bf 15}(5), S3.

\bibitem[Lauter {\em et~al.}(2014)Lauter, Lopez-Alt, and Naehrig]{Lauter2014}
Lauter, K., Lopez-Alt, A., and Naehrig, M. (2014).
\newblock Private computation on encrypted genomic data.
\newblock {\em Progress in Cryptology\/}.

\bibitem[Lippert {\em et~al.}(2017)Lippert, Sabatini, Maher, Kang, Lee, Arikan,
  Harley, Bernal, Garst, Lavrenko, Yocum, Wong, Zhu, Yang, Chang, Lu, Lee,
  Hicks, Ramakrishnan, Tang, Xie, Piper, Brewerton, Turpaz, Telenti, Roby, Och,
  and Venter]{Lippert2017}
Lippert, C., Sabatini, R., Maher, M.~C., Kang, E.~Y., Lee, S., Arikan, O.,
  Harley, A., Bernal, A., Garst, P., Lavrenko, V., Yocum, K., Wong, T., Zhu,
  M., Yang, W.-Y., Chang, C., Lu, T., Lee, C. W.~H., Hicks, B., Ramakrishnan,
  S., Tang, H., Xie, C., Piper, J., Brewerton, S., Turpaz, Y., Telenti, A.,
  Roby, R.~K., Och, F.~J., and Venter, J.~C. (2017).
\newblock {Identification of individuals by trait prediction using whole-genome
  sequencing data}.
\newblock {\em Proceedings of the National Academy of Sciences\/}, {\bf
  114}(38), 10166--10171.

\end{thebibliography}

\end{document}